\documentclass[aps,prl,showpacs, amsmath, amsfonts, amssymb, twocolumn]{revtex4}
\usepackage{graphicx}
\usepackage[usenames]{color}

\usepackage{ulem}  

\begin{document}

\title{Adler synchronization of spatial laser solitons pinned by defects}

\author{P.V.~Paulau$^1$, C.~McIntyre$^2$, Y. Noblet$^2$,
N. Radwell$^{2}$\thanks{Now at SUPA and Department of Physics and
Astronomy, University of Glasgow}, W.J.~Firth$^2$, P.~Colet$^3$,
T.~Ackemann$^2$, and G.-L.~Oppo$^2$}

\affiliation{$^{1}$ TU Berlin, Institut f\"ur Theoretische Physik,
Hardenbergstr. 36, Sekr EW 7-1, 10623 Berlin, Deutschland}
\affiliation{$^{2}$ SUPA and Department of Physics, University of
Strathclyde, 107 Rottenrow, Glasgow G4 0NG, UK} 
\affiliation{$^{3}$ IFISC, (CSIC-UIB), Campus Universitat Illes 
Balears, E-07071 Palma de Mallorca, Spain}

\pacs{42.65.Tg; 05.45.Xt; 05.45.Yv}

\begin{abstract}
Defects due to growth fluctuations in broad-area semiconductor 
lasers induce pinning and frequency shifts of spatial laser 
solitons. The effects of defects on the interaction of two solitons 
are considered in lasers with frequency-selective feedback both 
theoretically and experimentally. We demonstrate frequency and 
phase synchronization of paired laser solitons as their detuning 
is varied. In both theory and experiment the locking behavior 
is well described by the Adler model for the synchronization of 
coupled oscillators.
\end{abstract}

\maketitle

Spontaneous breaking of the translational symmetry in spatio-temporal
systems leads to the formation of nonlinear structures such as patterns,
solitons, oscillons, vortices and disorder \cite{Cross09}. Laser cavity 
solitons (LCS) are nonlinear self-localized dissipative states that 
possess both translational and phase invariance. The interaction of LCS 
leads to phase-locked bound states with well defined phases and separations 
as predicted for example in the cubic-quintic Complex Ginzburg Landau (CGL)
equations for temporal solitons in mode-locked lasers
\cite{Malomed91,Akhmediev97,Akhmediev02,
Leblond06,Turaev07,Soto07,Zavyalov09} and in models of lasers with
saturable absorbers for the spatial case
\cite{Vladimirov01,Rosanov02,Vahed11}. Corresponding bound
states have been observed experimentally in fiber lasers
\cite{Tang01,Grelu02}.

Spatial LCS have recently been observed in semiconductor-based
micro-resonators with either frequency-selective feedback
\cite{Tanguy08,Radwell09} or saturable absorption \cite{Genevet08,Elsass10}.
For temporal LCS, such as those arising in fiber lasers, the effects
of longitudinal inhomogeneities are washed out by the propagation
dynamics along the cavity axis, and every soliton sees the same material
characteristics. Spatial LCS in real systems are usually pinned by 
defects resulting from fluctuations during the epitaxial growth process
\cite{Radwell09,Genevet10}. Besides fixing the position, these defects
induce a shift in the LCS natural frequency. The frequency shift depends
on the characteristics of the defect itself and typically is different
for each of them. This diversity in natural frequencies is a critical
ingredient for the description of spatial LCS in real systems. Therefore,
despite being suitable for temporal LCS, theoretical studies considering
the interaction of identical LCS arising on a homogeneous background
are not adequate to describe the dynamics of coupled spatial LCS. Here
we show that the interaction of pinned LCS with different intrinsic
frequencies can be suitably described by the Adler locking mechanism
\cite{Adler46}. In particular we show that the coupling, if strong enough, 
can overcome the natural disorder leading to a synchronous regime
characterized by emission at a common frequency with a phase difference
that has a precise dependence on the frequency difference. The Adler
locking mechanism has relevance in biological clocks, chemical reactions,
mechanical and electrical oscillators \cite{Pikovsky01}. In optics
frequency locking of the Adler type was first observed in lasers 
with injected signals \cite{Buczek73} with more recent
generalizations to coupled lasers \cite{Fabiny93}, the spatio-temporal
domain \cite{Coullet98}, quantum dot lasers \cite{Goulding07} and
frequency without phase lockings \cite{Thevenin11}.

We first present frequency locking and phase synchronization of
spatial LCS pinned by defects in a general CGL model with
frequency-selective feedback where spatial variations of the cavity
tuning parameter are used to simulate the presence of background
defects. To show universality, defect induced Adler synchronization
is then demonstrated in a model closer to the experimental realization
where the saturable carrier dynamics are included \cite{Scroggie09}.
Finally, the phenomenon is demonstrated experimentally in a Vertical
Cavity Surface Emitting Laser (VCSEL) with an external Bragg grating
that provides frequency-selective feedback \cite{Radwell09}.

The interaction and locking phenomena which we observe in a
semiconductor laser with feedback are well captured in a simple
generic model consisting of a cubic CGL equation where solitons
are stabilized by coupling to a linear filter equation \cite{Paulau11}:
\begin{equation}
\left.
\begin{array}{l}
\partial_t E =g_{0}E+g_{2}|E|^{2}E - i \partial^2_x E+F + i n(x) E,
\\
\\
\partial_t F =-\lambda F+\sigma E\, ,
\end{array}%
\right.   \label{modeleqs}
\end{equation}%
where $E(x)$ is the intra-cavity field and $F(x)$ is the filtered
feedback field. Note that the linear feedback equation breaks the
Galilean invariance of the cubic CGL equation. For clarity reasons, we
focus here on one transverse spatial dimension. The time and space
coordinates ($t,x$) are scaled to 1~ns and 40~$\mu$m, respectively, 
so that $g_0$, describing linear gain and detuning, and $g_2$, 
describing nonlinear gain and dispersion, are dimensionless. We 
consider pure diffraction which is appropriate for VCSEL systems.
The real function $n(x)$ describes spatial variations of the cavity
tuning due to background defects that predominantly perturb the
material refractive index. In the second equation of (\ref{modeleqs})
$\sigma$ is the feedback strength, $\lambda$ its bandwidth, and we
have implicitly set our reference frequency to the peak of the filter
response. System (\ref{modeleqs}) has exact solutions corresponding
to stable single-frequency chirped-sech solitons \cite{Paulau11}.
Small variations of $n(x)$ lead to pinning and small changes
in the soliton frequency. The interaction of spatially separated
pinned solitons can lead to their locking which is our main interest
here.

We consider parameter values given by: $g_0=-4+28i$, $g_2=-96 -48i$,
$\lambda = 2.71$, $\sigma=162.6$, for which, in the ideal case with
translational invariance, $n(x)=0$, system (\ref{modeleqs}) has
stable solitons with two free parameters: location and phase. The
interaction of two such solitons makes them spiral slowly to fixed
relative distances $L$ and a phase difference $\Phi = \phi_2-\phi_1
= \pi/2$ unless merging takes place. $\Phi$ equal to zero and $\pi$
are also possible but correspond to saddles that are either phase or
distance unstable. Analytically the attainment of a bound state
reduces to the analysis of two transcendental equations in the
($L$,$\Phi$) phase space. The situation is very similar to that
described in \cite{Akhmediev97,Vladimirov01,Turaev07} for bound
solitons.

We now consider the case where the interaction takes place between
pinned solitons since the translational invariance is broken by the
pinning potential $n(x)$ which is equal to zero everywhere except
in the intervals $x_j-W<x<x_j+W$ where
\begin{equation}
n(x) = \frac{-n_j}{2} \; \left[ cos
\left( \frac{\pi (x-x_j)}{W} \right) + 1 \right]  \\
\end{equation}
with $j=1,2$. The pinning potential is a smooth function of $x$ and 
the width $2W$ of the defects is chosen to be close to the width at
half maximum of the LCS to help a quick convergence of the soliton
distance to the final defect separation.
Differences between the defects are described by the depths $n_1$
and $n_2$ of the pinning potential. The values of $n_j$ considered
here preserve the structure of the LCS with only its frequency
$\omega_j$ shifted.

\begin{figure}[htb]
\begin{center}
\includegraphics[width=8cm,keepaspectratio=true,clip=true]{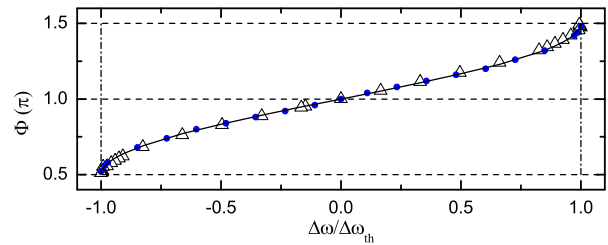}
\end{center}
\caption{(Color online) Locked phase differences $\Phi$ of pinned LCS
for different frequency detunings (controlled by the potential depths
$n_1$ and $n_2$) from direct integration of Eq.~(\ref{modeleqs})
(dots, LCS separation of 5.3 soliton widths) and the model of
Ref.~\cite{Scroggie09} (triangles, LCS separation of 4 soliton widths).
The solid line refers to the Adler equation (\ref{Adler}).}
\label{Fig1}
\end{figure}

If the defects are close enough in space, the soliton interaction 
locks their frequencies and phases to common values that depend on 
the difference between the defect depths. The synchronization dynamics 
of the phase difference $\Phi$ between the pinned solitons relaxes 
to well determined stationary values that depends on the defect detuning 
parameter $\Delta \omega = \omega_2 - \omega_1$ generated by the choice 
of $n_1$ and $n_2$ values. The dependence of the stationary phase
difference $\Phi$ on the detuning $\Delta \omega$ for numerical
simulations of (\ref{modeleqs}) is shown in Fig.~\ref{Fig1} for
$|x_2-x_1|=1.5$ space units. There are a maximal and a minimal
detuning $\pm \Delta \omega_{th}$ below and above which synchronization
does not take place. Very similar results have been obtained from
numerical simulations of LCS in models of VCSELs with frequency-selective
feedback that include the dynamics of the carriers and more realistic
values of the linewidth enhancement factor \cite{Scroggie09}
(see Fig.~\ref{Fig1}).

The archetypical equation describing synchronization between two coupled
oscillators is the Adler equation \cite{Adler46},
\begin{equation}
\frac{d \Phi}{dt} = \Delta \omega - \varepsilon \sin(\Phi),
\label{Adler}
\end{equation}
where in-phase and anti-phase solutions are selected for zero detuning,
$\Delta \omega =0$, depending on the sign of the coupling parameter
$\varepsilon$: for positive $\varepsilon$ the final stable state is
$\Phi=0$; for negative $\varepsilon$ it is $\Phi=\pi$. A comparison of
the results of the Adler equation with negative $\varepsilon$ and the
simulations of the synchronization of LCS in both equations
(\ref{modeleqs}) and the model of Ref. \cite{Scroggie09} is presented
in Fig.~\ref{Fig1}. The agreement is remarkable. Note that the $\pi/2$
value observed in phase locking of dissipative solitons without defects
\cite{Grelu02} is now replaced by the $\pi$ value typical of Adler
synchronization. In-phase and out-of-phase values have already been
observed in numerical simulations of LCS in cubic-quintic CGL equations
with regular variations of the background \cite{He09,Chang09} although
no Adler scenario is suggested. In particular unless the period of the
modulation is much larger than the length scales due to soliton interaction,
the LCS are forced into different minima of the potential and do not
experience any detuning difference anymore \cite{Chang09}. This is
consistent with the $\pi$-phase states we observed for localized defects
of equal depths ($|n_2-n_1|=0$).

\begin{figure}[htb]
\begin{center}
\includegraphics[width=8cm,
keepaspectratio=true,clip=true]{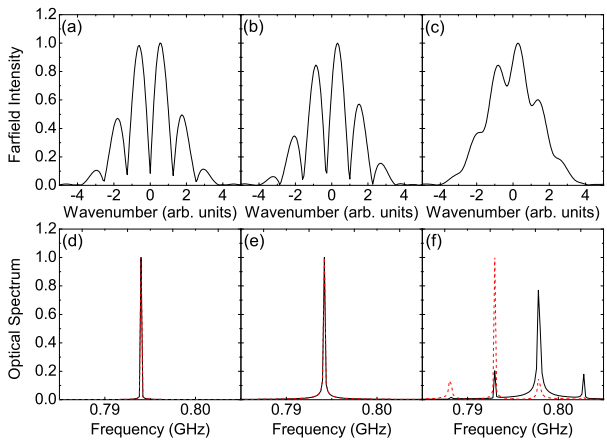}
\end{center}
\caption{(Color online) Far field fringes (a)-(c) averaged over 2~$\mu$s,
and optical spectra (d)-(f) for a time window of 5~$\mu$s, for
$\Delta \omega / \Delta \omega_{th} =0$ (a,d), $0.99$ (b,e) and $2.0$
(c,f) obtained from simulations of the model of Ref.~\cite{Scroggie09}.
In (d,e) the LCS spectral peaks (dashed and solid lines) overlap.}
\label{Fig2}
\end{figure}

To characterize the Adler locking both in the spatial and temporal
domains, we display the time averaged far field images in the top
part of Fig.~\ref{Fig2} and the optical spectra in bottom part of
Fig.~\ref{Fig2} for two points inside ($\Delta \omega / \Delta
\omega_{th} =0$ and $0.99$) and one outside the Adler region ($\Delta
\omega / \Delta \omega_{th}=2$), respectively. Progressive change
of the LCS phase difference $\Phi$ (from $\pi$ in Fig.~\ref{Fig2}a
to around $1.5\pi$ in Fig.~\ref{Fig2}b) is reflected in the change
in the symmetry of the fringe pattern. Far field fringes are
well defined in the region where the LCS are locked in frequency
(see the full overlap of the soliton peaks in the frequency spectrum
in Fig.~\ref{Fig2}d and \ref{Fig2}e) indicating a strong interaction.
For detunings much larger than the locking range, the fringe visibility
disappears and the spectrum is formed just by the lines of the
individual solitons (not shown) corresponding to LCS operating
independently. For detunings just outside the Adler locking region,
however, some phase and spectral correlation survives due to 
non-uniform evolution of the relative phase (Figs.~\ref{Fig2}c 
and \ref{Fig2}f).

The Adler locked state between LCS is a robust feature independent of
initial conditions such as initial phases, frequencies and sequential
order of creation of the two LCS. Once the locked state is attained,
one of the two LCS can be switched off by a short, localized perturbation
to the carrier density at its location. Hence, LCS retain their solitonic
properties in the phase-locked state in the sense that they are still
individually bistable and optically controllable.

\begin{figure}[htb]
\begin{center}
\includegraphics[width=8cm,
keepaspectratio=true,clip=true]{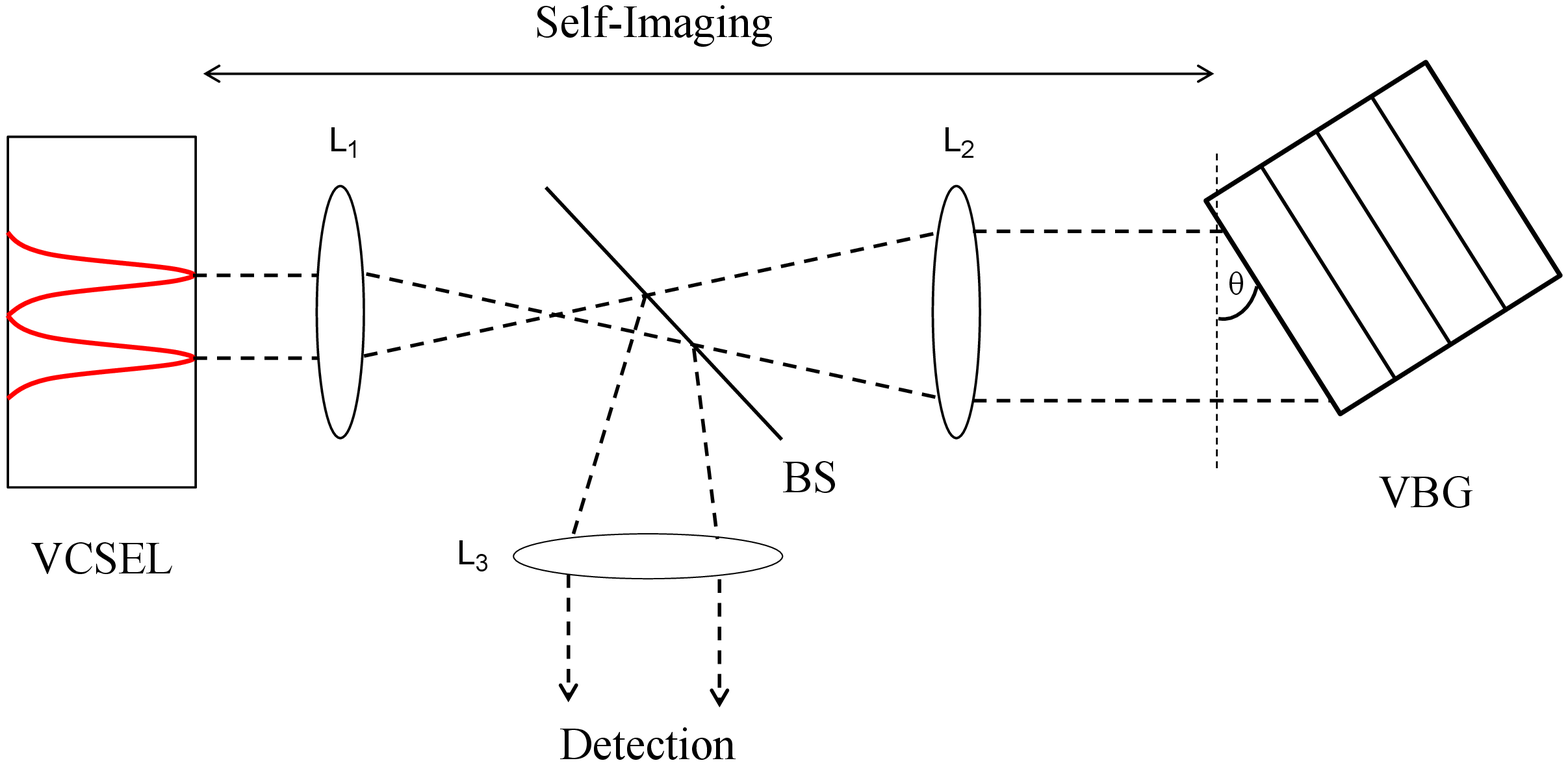}
\end{center}
\caption{Schematic diagram of a self-imaging cavity coupling a VCSEL
to a frequency-selective element. The beam profiles indicate two 
interacting LCS. The dotted lines correspond to the centers of the 
bundle of rays emitted by the soliton. The tilt angle $\Theta$ is 
greatly exaggerated for clarity of display. The focal lengths of the
intra-cavity lenses L$_1$, L$_2$ are $f_1=8$~mm and $f_2=50$~mm, 
the total cavity length $L \approx~12$~cm with a 1.23~GHz free 
spectral range. L$_3$ images the near field of the VCSEL into the 
detection arm.}
\label{Setup}
\end{figure}

The experiment has been performed with a temperature tuned 981~nm
VCSEL of 200~$\mu$m circular aperture and a volume Bragg grating
(VBG) with a single reflection peak at 981.1~nm, a reflection
bandwidth of 0.2~nm full-width at half-maximum (FWHM) and a peak
reflectivity of 99\% \cite{Radwell09}.
The external cavity for the frequency-selective feedback is arranged
in a self-imaging configuration that maintains the high Fresnel
number of the VCSEL cavity and ensures local feedback compatible
with self-localization (see Fig. \ref{Setup}). Small deviations from
the self-imaging condition are not critical for the reported phenomena.
The detection system comprises two
charge-coupled-device cameras for near- and far-field imaging, and a
scanning Fabry-Perot interferometer with a 10~GHz free spectral range
to measure the optical spectrum. Several LCS
appear at certain spatial locations defined by the traps when
increasing the VCSEL injection current and display hysteretic
behavior when decreasing the current again. The experiment described
below is performed at a bias current at which both LCS involved are
individually bistable. Investigations were performed on pairs of
different LCS with a distance of 30 to 80~$\mu$m. We focus here on a
configuration of two LCS a distance of 79~$\mu$m, but the results
are typical also for the other configurations. Each of these LCS is
a coherent emitter but they are usually mutually incoherent due to
the disorder \cite{Radwell09,Genevet10}.

\begin{figure}[htb]
\begin{center}
\includegraphics[width=8cm,
keepaspectratio=true,clip=true]{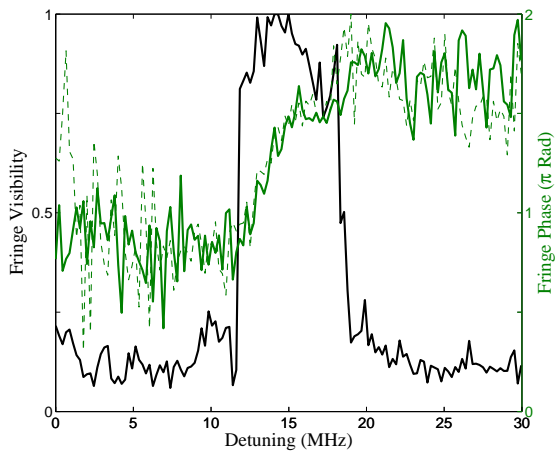}
\end{center}
\caption{(Color online) Fringe visibility (black) and fringe phase
(green curves, gray in print) as a function of the tilt angle that
changes the difference between the feedback phases of the LCS.
This difference is converted to a frequency scale by multiplying
it by the free spectral range of the external cavity thus providing 
the change of the relative detuning between the two LCS in the
external cavity. The zero of this detuning scale is arbitrary. The
solid and dashed green curves are obtained for scanning the tilt
back and forth. The fringe phase is obtained from the phase of a
cosine-wave fitted to far field profiles like those in the upper row
of Fig.~\ref{Fig4}. Other parameters: Temperature $69^\circ$C,
current $I= 373$~mA.}
\label{Fig3}
\end{figure}

Since it is experimentally awkward to vary the detuning between
two LCS by locally changing the properties of the VCSEL itself,
we use a piezo-electric transducer to minutely tilt the external
cavity's end reflector (VBG) with respect to the optical axis. This
leads to a differential change of the external cavity length for
the two LCS and thus to a differential change in feedback phase,
which can be incorporated into Eqs.~(\ref{modeleqs}) by making
$\sigma$ complex. In this way the frequency difference, i.e.\ the
detuning $\Delta \omega$, between two LCS can be tuned \cite{Microchip}.
During the scan, LCS position in near field and angular center in
far field stay constant to better than 5\% and 2.5\% of their width, 
respectively. When performing such a scan, a region of frequency
and phase locking appears, identified in Fig.~\ref{Fig3} by the
region of high fringe visibility in the far field. These
fringes are video integrated over a time of 20~ms
(significantly longer than any intrinsic time scale) and last for
seconds to hours depending on parameters. This illustrates that
locking -- once achieved by a careful alignment of the VBG -- is
a robust phenomenon.

As expected for the Adler scenario, in the locking region, the fringe
phase varies smoothly and quasi-linearly with the detuning of the
external cavity. It is much more noisy outside, where the visibility
is low. There is no significant phase hysteresis when the tilting is
reversed (see the green solid and dashed lines in Fig.~\ref{Fig3}),
again as expected for the Adler scenario. The transitions to and from
frequency and phase-locking are rather abrupt (Fig.~\ref{Fig3}, black
curve). For clarity, we show only a single sweep of the fringe visibility,
because there is significant jitter at the transition points. This and
the fact that the locking range is only about $\pi/2$ can be attributed 
to features beyond the phase-only approximation underlying the Adler
equation such as the multi-longitudinal mode structure and possibly 
amplitude dynamics \cite{Wieczorek05}. 
Longitudinal mode hopping of individual solitons can enable and/or 
quench the Adler dynamics thus explaining the jitter and limited 
locking range of (Fig.~\ref{Fig3}). Within the locked region, however, 
the dynamics follows the Adler scenario with the locking phase being 
determined by the solitons' differential feedback phase.

The importance of the external cavity structure is also evidenced by
the fact that locking-unlocking scenarios can be induced by changing
the VCSEL current if the VBG is adjusted close to the locking region.
In this case the refractive index effects due to the
ohmic heating shift the cavity resonances and induce
hopping between different external cavity modes for each
individual soliton. The LCS can then lock to a common
external cavity mode for some range of the injection current 
and display again Adler interaction.

\begin{figure}[htbp]
\begin{center}
\includegraphics[width=8cm,keepaspectratio=true,clip=true]{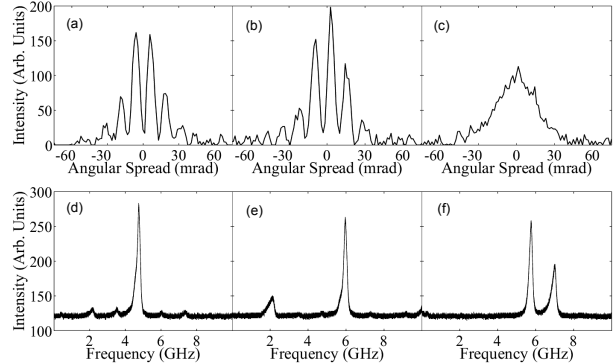}
\end{center}
\caption{Upper row: cut through far field intensity distribution
orthogonal to fringe orientation. Lower row: optical power spectra.
Left column (a,d) for detunings around 12~MHz, locked with a phase
of~$\pi$; center column (b,e) around 18 MHz, near the end of the
locking region, locked with a phase of 1.5~$\pi$; right column
(c,f) around 22 MHz, unlocked, no clear fringes.}
\label{Fig4}
\end{figure}

Fig.~\ref{Fig4} shows experimental far-field fringes (upper
part) and the corresponding optical power spectra (lower part), to
be compared with the numerical results of Fig.~\ref{Fig2}. When the
fringe visibility is high (Figs.~\ref{Fig4}a, b), the two LCS have the
same frequency (Figs.~\ref{Fig4}d, e). Weak sidemodes indicate some
residual excitation of neighboring external cavity modes. The change
in fringe phase from~$\pi$  (Fig.~\ref{Fig4}a) to 1.5~$\pi$
(Fig.~\ref{Fig4}b) is reflected in the change in symmetry of the
fringe pattern. Outside the locking region the fringes essentially
disappear (Fig.~\ref{Fig4}c) and the two LCS operate on different
frequencies. 

Synchronization behavior has been discussed in both
continuous and coupled oscillator models \cite{Pikovsky01}.
Our study uses a continuous model, but synchronization is
between `discrete' entities, the solitons. As such, self-localized
solitonic oscillators provide a nice bridge between spatially
extended media and coupled, pre-defined oscillators. Although we
have demonstrated the validity of Adler's model for just two
solitons, we suggest that network synchronization in the spirit 
of Kuramoto's model (with coupling possibly controlled by the
deviation from the self-imaging condition) should be possible with
many LCS in a fruitful analogy with brain activity \cite{Cumin07}
and, possibly, with spatio-temporal excitability \cite{Coullet98}.

In conclusion, we have demonstrated spatio-temporal Adler
synchronization without injection in semiconductor lasers with
frequency selective feedback. The synchronization is induced by
spatial defects where the LCS are pinned. The presence of the
defects breaks the translational symmetry, fixes the relative
distance between solitons and locks the relative phase to values
different from $\pi/2$ observed numerically in the absence of
defects or experimentally in temporal-longitudinal systems.
A regime of Adler synchronization is identified when changing
the frequency of each soliton with respect to that of its neighbour.

P.V.P. acknowledges support from SFB 910; C.M., Y.N., and
N.R. from EPSRC DTA; P.C. from MICINN and Feder (FIS2007-60327, 
FISICOS, TEC2009-14101, DeCoDicA).


\end{document}